\documentclass[10pt]{article}
\usepackage{verbatim}
\usepackage{apalike}
\usepackage{doublespace}
\usepackage{fancyheadings}
\usepackage{graphics}
\usepackage{epsfig}

\begin{document}
\begin{spacing}{1}
\title{
Coding strategies in monkey V1 and inferior temporal cortices}
\date{}
\author{Ethan D. Gershon\thanks{
        current address: New York University
                        School of Medicine and Center for Neural Science,
                        New York, NY 10016},\\
	Matthew C. Wiener,\\
        Peter E. Latham,\thanks{current address:  Dept. of Neurobiology, UCLA, Los Angeles, CA, 90095.} \\
        and Barry J. Richmond;\thanks{
                        reprint requests and correspondence:
                        Laboratory of Neuropsychology,
                        49 Convent Drive,
                        Bethesda, MD 20892-4415,
                        Phone: (301) 496-5625 x225,
                        fax: (301) 402-0046,
                        E-mail: bjr@ln.nimh.nih.gov}\\
\\
\\
Laboratory of Neuropsychology,\\
National Institute of Mental Health,\\
and Laboratory of Developmental Neurobiology,\\
National Institute of Child Health and Human Development,\\
National Institutes of Health, Bethesda, MD 20892.
 \\
 \\
Short title:  Coding strategies in monkey V1 and IT cortices. \\
\\
\\
\\
\\
\\
published version:  J. Neurophysiol., 79: 1135-1144, 1998 \\
}
\maketitle

\end{spacing}
\begin{spacing}{1}
{\bf Abstract}
We would like to know whether the statistics of neuronal
responses vary across cortical areas.
We examined stimulus-elicited
spike count response distributions in V1 and IT cortices of awake monkeys.
In both areas the distribution of spike
counts for each stimulus was well-described by a Gaussian,
with the log of the variance in the spike count linearly related
to the log of the mean spike count.
Two significant differences in
response characteristics were found:  both the range of spike counts
and the slope of the log(variance) vs. log(mean) regression were larger
in V1 than in IT.   However, neurons in the two areas transmitted approximately
the same amount of information about the stimuli, and had about the
same channel capacity (the maximum possible transmitted information given
noise in the responses).
These results suggest that neurons in V1 use more variable signals
over a larger dynamic range than neurons in IT, which use less
variable signals over a smaller dynamic range.  The two coding strategies
are approximately as effective in transmitting information.

\section*{INTRODUCTION.}

Neurons in different regions of the
visual system encode different aspects of visual stimuli.
For example, neurons in V1 cortex respond
strongly to an oriented bar while those in inferior temporal (IT)
cortex often require a more complex stimulus.
We would like to know whether the differences in what is encoded
are reflected in differences in the neuronal response,
as this may shed light on strategies for cortical processing.
Specifically, we ask two questions.  First, in what way does the
statistical structure of responses
differ across areas?  
Second, 
how do the differences, if any, affect information transmission?

The first question can be answered
by looking at the statistics of neuronal responses.
Depending on the nature of those responses,
the relevant statistics may be as simple as mean firing rate or as
complicated as high order correlations in spike arrival times.
The second question requires a precise definition of information,
which is provided by  
information theory.   Information theory tells us that
the ability of a neuron to distinguish among members of a
set of stimuli 
depends on two things.  The first is the number of distinguishable
responses:  the more distinguishable responses, the larger the information.
The second is the variability in the response to each stimulus.
Variability (which is reflected in the
probabilistic transformation from stimulus to response)
clearly degrades information transmission.  
If each stimulus produces a very broad range of responses, the
information in each response may be very small no matter how many
distinguishable responses there are.

A difficulty in applying information theory to neuronal systems
is determining what to call a response, i.e., what coding scheme to use.
Two extremes are the number of spikes in a fairly wide time window, and
the spike arrival times measured with high resolution.
In general one looks for the simplest coding scheme that conveys the
most information | conflicting constraints whose relative importance
must be decided on a case by case basis.
Here we have the additional problem that we want to compare
brain regions that may use different coding schemes.
Fortunately, for V1 and IT, the areas we consider here,
it has been shown that spike count in a window approximately
300 ms wide carries most of the stimulus-related
information | about 80\% \cite{Heller95}.  The remaining 20\% of the
information is carried in spike timing with an accuracy of about 30 ms
in V1 and 60 ms in IT \cite{Heller95}.  Since neurons in V1 fire at about twice
the rate as those in IT (see Results), the spike timing accuracy relative
to the mean interspike interval is about the same in the two areas.
Thus, in this paper we use the spike count as our neural code.  
This assumption greatly simplifies our calculations,
although in principle everything we do here could be applied to
coding schemes that include temporal variations.

Since we are using a spike count code, the number of distinguishable responses
is the range of spike counts a neuron is capable of producing in response
to {\it all} stimuli; we refer to this as the dynamic range.
Any {\it single} stimulus also elicits a range of spike counts.
This variability in the response to a single stimulus is captured in the
conditional probability of observing $n$ spikes given stimulus $s$, $P(n|s)$.
Therefore, we can 
compare responses in the two regions by examining only the dynamic
range and the conditional probabilities.
In practice, this means we need a stimulus set only large enough to provide
accurate estimates of these two quantities.  This is a
weaker constraint on the stimulus set than is required by other
information theoretic analyses, which
depend on the frequencies with which stimuli are presented
\cite{CoverThomas}.

Determining the dynamic range of a neuron is relatively 
straightforward; the
difficult part of the analysis is determining the
conditional probability distribution, $P(n|s)$.  Here we follow
previous work, in which it has been shown that the logarithm of
the variance of the stimulus-elicited spike count is linearly
related to the logarithm
of the mean spike count in both cat and monkey V1 cortex
\cite{Tolhurst81,Dean81,Tolhurst83,vanKan85,Vogels89}.  We confirm
the mean-variance relation in monkey V1, and we observe a similar
relation in monkey IT cortex.  We then go on to show that
$P(n|s)$ is well approximated by a modified Gaussian distribution
(the main modification was truncation at 0; see Methods for details) 
with mean,
$\mu$, that depends on the stimulus, $s$, and variance that depends only
on the mean.

What the mean-variance relation gives us is the conditional probability,
$P(n|\mu)$, of observing $n$ spikes given a mean spike count $\mu$.
While $P(n|\mu)$ is not quite the same as the probability of observing $n$
spikes given the {\it stimulus}, it is in some ways more valuable.
For example, given the mean-variance relation,
and thus $P(n|\mu)$, in V1 and IT, it is relatively easy to
compare the two areas:  simply examine which area
has a larger variance for a given mean spike count, which has a larger
dynamic range, and investigate how the differences
affect information transmission.
When we carry out this exercise,
we find that both the spike count variance
and the dynamic range are significantly larger in V1 than in IT.
For information transmission, these two trends work
in opposite directions:
a large dynamic range increases information transmission (more
distinguishable responses) while a large variance reduces information
transmission (more variability).  For V1 and IT
the two effects approximately cancel:  the maximum amount of information
that could be transmitted is about the same in the two areas | assuming
a spike count code is used, and the observed
dynamic range and mean-variance relations apply.
Thus, although neurons in V1 and IT implement different coding strategies,
as reflected in the significantly
larger variability and range of responses in V1 than in IT,
neurons in the two areas are capable of transmitting about
the same amount of information using a spike count code.

An abstract of these results has appeared \cite{Gershon96}.

\section*{METHODS}

\subsection*{Data set.}

We performed new analyses using
previously published data.
The data came from two studies of supragranular
V1 complex cells, each study using 
two rhesus monkeys performing a simple fixation
task \cite{bjr90I,Kjaer97}, 
and from one study of neurons in area TE of IT cortex in two other monkeys
performing a simple sequential nonmatch-to-sample task
\cite{Eskandar92I}.
The stimuli were centered on V1 neuronal receptive fields
which were located in the lower contralateral visual field
one to three degrees from the fovea.  The IT visual receptive fields
were large, bilateral and included the fovea.  Standard
extracellular recording methods were used throughout.

\begin{figure}
\epsfig{file=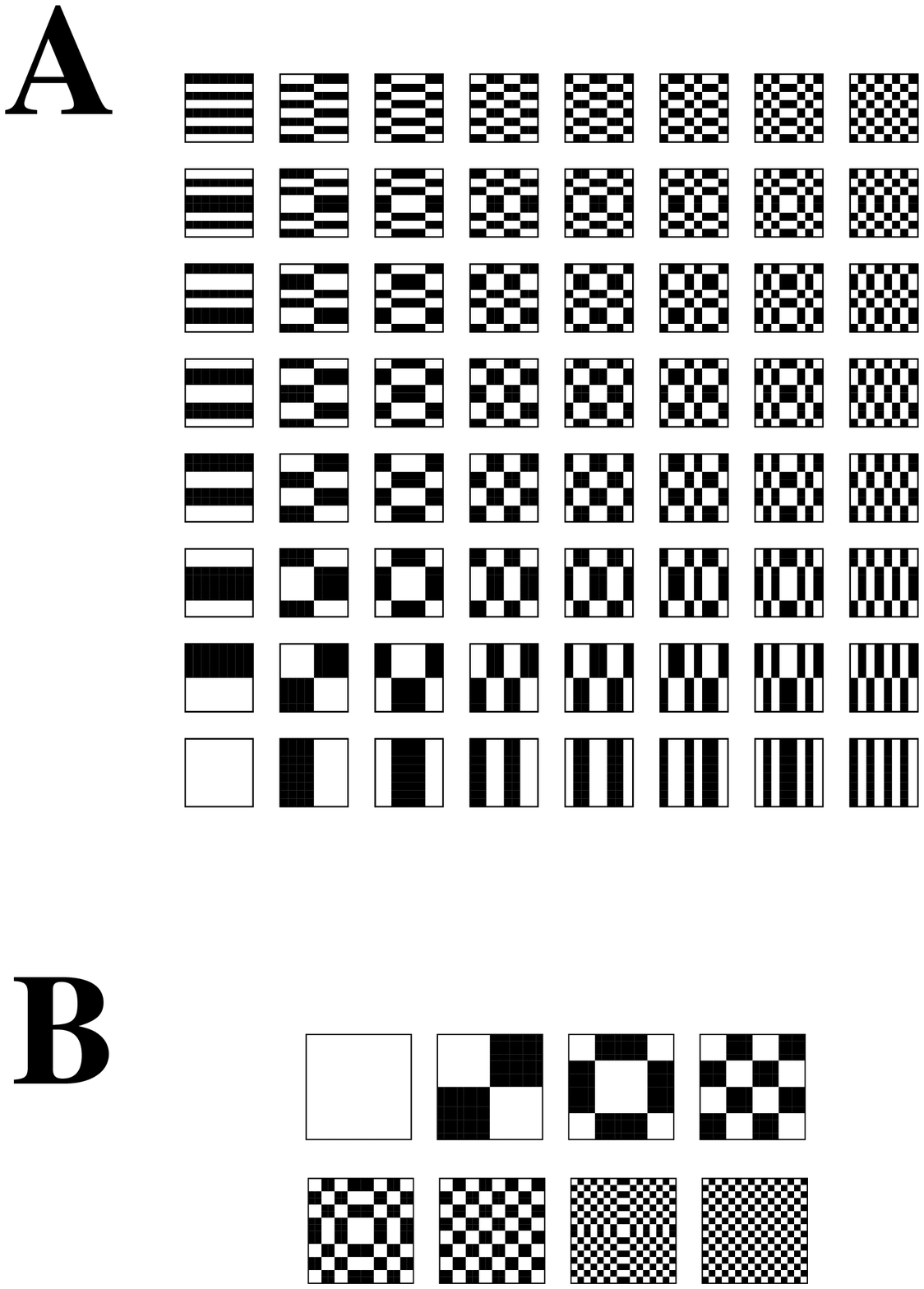,height=3.5 in,width=3 in}
\caption{The Walsh patterns.  For V1 set 1, 
the 64 stimuli labelled A and the corresponding
contrast-reversed set were presented on the receptive fields while the
monkeys fixated.  The stimuli were $2.5^\circ$ on a side (covering the
excitatory receptive field and some of the surround).  For V1 set 2, the 8
stimuli labelled B and the corresponding contrast-reversed set
were presented on the receptive field while the monkey fixated.
For IT, the 4
$\times$ 4 set (16 stimuli) in the lower left corner of set A and the
corresponding contrast-reversed set were used as the monkey performed a
non-match-to-sample task.  The stimuli were $4^\circ$ on a
side and were centered at the point of fixation.
}
\label{walsh.fig}
\end{figure}

In the three experimental studies considered here
the visual stimuli were two
dimensional black and white patterns based on the Walsh functions
(Figure \ref{walsh.fig}).  
For V1 set 1 \cite{bjr90I}, 128 stimuli were used:  
a set of sixty-four $8 \times 8$ pixel patterns and
their contrast-reversed counterparts.  For
V1 set 2 \cite{Kjaer97}, 16 stimuli were used:
a set of eight $16 \times 16$ pixel patterns
and their contrast-reversed counterparts.
In both sets, the patterns covered the excitatory receptive field.
At $3^{\circ}$ eccentricity, the stimuli were about $2.5^{\circ}$
on a side. 
For the IT
experiments 32 stimuli were used:  
16 $4 \times 4$ pixel patterns and their
contrast-reversed counterparts.  
These patterns were $4^{\circ}$ square, 
and centered on the fixation point.
 
The stimulus was on for 320 ms in V1 and 352 ms in IT.
To account for latencies and to avoid contamination from off-responses,
spikes were counted during
the interval from 30-300 ms after stimulus onset for the V1 neurons
and 50-350 ms after stimulus onset for the IT neurons.  For every
neuron, each stimulus was presented approximately the same number of
times ($\pm 2$), in randomized order.  
Different neurons received different numbers of
presentations.  The number of stimulus presentations was between 4 and
50 in V1, and between 19 and 50 in IT.  The timing of events,
including spikes, was recorded with 1 ms resolution.

\subsection*{The relationship between mean spike count and its variance.}

For each cell, each stimulus produces a sample mean spike count,
$\mu_i$, and a sample variance in spike count, $\sigma^2_i$, 
where the subscript $i$ labels stimulus.
We use
linear regression to fit the curve $\log \sigma^2 = b + m \log \mu$
to the set of points $(\mu_i, \sigma_i^2)$.
This results in a slope, $m$, and intercept, $b$, for each cell.

Estimates of $\log(\mu)$ and $\log(\sigma^{2})$ obtained by 
taking the logarithm of the sample mean and variance are 
biased and result in underestimation of the variance 
of response distributions and overestimation of transmitted 
information.  We corrected for the bias
using a Taylor series expansion;
only a few terms are needed for good results.  
See Kendall and Stuart, 1961, pp. 4-6.

\subsection*{Fitting analytic distributions to the data.}

We seek a
model for the conditional probability, $P(n|s)$, of observing $n$
spikes in response to stimulus $s$.  We 
examined two widely-used probability distributions -- the Poisson 
distribution and a modified Gaussian distribution.  
The Gaussian distribution was modifed by truncation to eliminate
the negative portion, followed by normalization.
Such distributions have been considered for neural data before
\cite{Foldiak93}.  
The probability of
seeing $n$ spikes was taken to be the integral of this density
function between $n-\frac{1}{2}$ and $n+\frac{1}{2}$ 
($0$ and $\frac{1}{2}$ for $n=0$).
A $\chi^{2}$ test was used to compare each of the analytic 
distributions to the histogram of experimentally observed 
spike counts.
To have enough data for this analysis, 
only the responses to stimuli that had been presented 12 or more times 
to a given cell were considered.  

As an alternative to integrating the Gaussian between
$n-\frac{1}{2}$ and $n + \frac{1}{2}$, we took the probability
of observing $n$ spikes, $P(n|s)$, to be
proportional to the Gaussian density evaluated at $n$.
The constant of proportionality was chosen to ensure that the total
probability summed to one.  This alternative method resulted in negligible
differences in all quantities we calculated.

\subsection*{Information measures.}
 
The information carried in a neuron's response
about which member of a set of
stimuli is present is defined as \cite{CoverThomas}
 
\begin{equation}
I(S;R) = \sum_{s,r} P(s) P(r|s) \log_2 { P(r|s) \over P(r)}
\label{info.definition}
\end{equation}
 
\noindent
where $S$ is the set of stimuli $s$, $R$ is the set of responses $r$,
$P(r|s)$ is the conditional probability of response $r$ given
stimulus $s$, $P(s)$ is the probability that stimulus $s$ occurred,
and $P(r) = \sum_{s} P(r|s)P(s)$ is the probability of response $r$.  Equation \ref{info.definition}
is
general, but here we confine ourselves to the case where the response $r$
is taken to be the number of spikes elicited by the stimulus.  Thus, in
what follows we replace $P(r|s)$ with $P(n|s)$ and $P(r)$ with $P(n)$
where $n$ is the number of spikes.

The transmitted information, $I(S;R)$, given in equation 
\ref{info.definition} is a
function of the stimulus probability distribution, $P(s)$.
The channel capacity is the maximum value of $I(S;R)$ with
respect to the probability distribution $P(s)$.  Here we take $S$ to
contain all visual stimuli.  
Clearly channel
capacity depends on what we take for the response, i.e., what we choose
for the neural code.
However, once we choose a code and a stimulus set, the channel capacity
is well-defined and it represents a lower bound on the maximum amount of
information that could be transmitted.
In this analysis we use a spike count code.  Such a code
has been shown to carry about 80\% of the stimulus related
information \cite{Heller95}, so we suspect that the lower bound we compute
will not be far from the true maximum transmitted information.

Because the channel capacity is independent of the frequencies with
which stimuli are presented in any single experiment, 
it is a robust measure that can be used
to compare information transmission rates
across brain regions.  However, it is more difficult to compute than
transmitted information for purely experimental reasons:
we can measure the conditional probability distribution,
$P(n|s)$, for a relatively small number of stimuli, $s$, but
to accurately estimate the channel capacity we need to know
$P(n|s)$ for all stimuli.  We can get around this problem by first
constructing $P(n|\mu)$ from $P(n|s)$, where $P(n|\mu)$ is
the probability of observing spike count
$n$ given the mean spike count $\mu$, and  second, developing an
analytic model for $P(n|\mu)$.

The expression for transmitted information must be rewritten
in terms of $P(n|\mu)$.
We start by writing the transmitted information, $I(S;R)$, in the form

\begin{equation}
I(S;R) = \sum_\mu \sum_{s \in [\mu, \mu + \Delta\mu],n}
P(s) P(n|s) \log_2 { P(n|s) \over P(n)}
\label{reorder.stimuli}
\end{equation}

\noindent
where the notation $s \in [\mu, \mu + \Delta\mu]$ 
means restrict $s$ to only those
stimuli that produce a response whose mean spike count lies between
$\mu$ and $\mu + \Delta\mu$, and the sum over $\mu$ 
runs in increments of $\Delta\mu$.
Equation \ref{reorder.stimuli} is exact; all we have done
is order stimuli by the mean spike count they produce.
The next step is to replace
$P(n|s \in [\mu, \mu + \Delta\mu])$ with $P(n|\mu)$.
This would also be exact in the limit $\Delta\mu \rightarrow 0$
if the distribution of spike counts depended
only on the mean.  We show in the results
that it is a good approximation to assume that the distribution of
spike counts does depend only on the mean; in particular, it provides an
estimate of the transmitted information that is consistent with estimates
reached by other accepted methods.  Thus, we will adopt
that approximation here.

Before replacing
$P(n|s \in [\mu, \mu + \Delta\mu])$ with $P(n|\mu)$, we need to 
express $P(\mu)$ in terms of $P(s)$.
This can be done by noting that
$P(s)$ induces a probability distribution $P(\mu)$,

\begin{equation}
P(\mu) \, \Delta\mu = \sum_{s \in [\mu, \mu + \Delta\mu]} P(s) .
\label{prob.of.mean}
\end{equation}

\noindent
Then, ignoring the error associated with
the approximation $P(n|s \in [\mu, \mu + \Delta\mu]) \approx P(n|\mu)$,
we write the probability of observing spike count $n$, averaged
over all mean spike counts, as

\begin{equation}
P(n) = \int d\mu \, P(n|\mu) P(\mu)
\label{prob.spike.count}
\end{equation}

\noindent
where we replaced the sum over $\mu$ that appeared in equation 
\ref{reorder.stimuli} with
an integral, valid in the limit of small $\Delta\mu$.
Finally, we can rewrite equation \ref{reorder.stimuli} for $I(S;R)$ in
terms of probability distributions over $n$ and $\mu$,

\begin{equation}
I(S;R) = \int d\mu \ P(\mu) \sum_n
P(n|\mu) \log_2 { P(n|\mu) \over P(n)}
\label{info.from.mean}
\end{equation}

\noindent
with $P(\mu)$ and $P(n)$ given in equations 
\ref{prob.of.mean} and \ref{prob.spike.count}, respectively.
Again we use an integral over $\mu$ rather than a sum.
 
We show in the results that $P(n|\mu)$ is well approximated by a modified
Gaussian distribution whose variance is a function of mean spike count.
Using this modified Gaussian
we can determine channel capacity by finding the distribution
of mean spike counts, $P(\mu)$, that maximizes transmitted information,
equation \ref{info.from.mean}.  
That distribution must be found numerically, and 
the numerical implementation requires that we discretize
the continuous space of mean responses.  We denote these discretized
probabilities by $\bar{P}(\mu) = \int_{\mu}^{\mu+\Delta\mu} d\mu P(\mu)$.

The search for the maximizing 
set of probabilities is subject to three constraints:
(1)  the probabilities must be non-negative; (2) the probabilities must
sum to one; and (3) the range of means must be finite.  
The first two constraints arise from intrinsic properties of probability 
distributions.  If the third constraint is violated,
the transmitted information can be infinite and the problem of
maximizing transmitted information is ill-posed. 

The first constraint is implemented by restricting the search space such
that \mbox{$ 0 \le \bar{P}(\mu)$} for all $\mu$.
The second constraint is implemented by requiring that
\begin{equation}
\sum_{\mu} \bar{P}(\mu) = 1 \ \ .
\end{equation}
The third constraint
is implemented by requiring that the distribution of spike counts
be consistent with the observed data; that is, the distribution
of means
 must not lead to a distribution of spike counts
with many counts outside the observed range.
Specifically, if $n_{\min}$ and $n_{\max}$ are the minimum and maximum
observed spike counts over all stimuli for a particular cell, then we
demand that

\begin{equation}
\sum_{n > n_{\max}} C_+(n-n_{\max}) P(n) +
\sum_{n < n_{\min}} C_-(n_{\min}-n) P(n) = \epsilon
\label{G1}
\end{equation}

\noindent
where $P(n)$ is defined in equation \ref{prob.spike.count},
both $C_+(n)$ and $C_-(n)$ are non-decreasing functions of $n$,
and $\epsilon$ is small.  
Equation \ref{G1} 
ensures that $P(n)$ falls off rapidly for spike counts outside
the observed range.  To implement the optimization procedure we need to
translate this into a constraint on $\bar{P}(\mu)$, since the search for the
maximum value of the transmitted information occurs in $\bar{P}(\mu)$ space.
Defining the function

\begin{equation}
C(\mu) \equiv
\sum_{n > n_{\max}} C_+(n-n_{\max}) P(n|\mu) +
\sum_{n < n_{\min}} C_-(n_{\min}-n) P(n|\mu)
\label{Cn}
\end{equation}

\noindent
and combining equations \ref{prob.spike.count} and \ref{G1}, we arrive at

\begin{equation}
\sum_{\mu} \, C(\mu) \bar{P}(\mu) \leq \epsilon \ \ .
\label{constraint.3}
\end{equation}

\noindent

\noindent
Equation \ref{constraint.3} 
represents our third constraint.  
In practice, since expanding the
range of spike counts increases transmitted information, we do not
have to worry about our range being too small,
only too large.  Therefore, in equation \ref{constraint.3} only
the equality constraint is important.

In our numerical
calculations we use
\begin{eqnarray}
\nonumber
C_+(n) & = & C_-(n) = n^2 \ ,
\\
\nonumber
\epsilon & = & 0.1
\, .
\end{eqnarray}
\noindent
To find the channel capacity, we minimize the function
\begin{equation}
F[\bar{P}(\mu)] = - I(R;S) + h_{1} (\sum_{\mu} \bar{P}(\mu) -1)^2 +
h_{2} (\sum_{\mu} C(\mu)\bar{P}(\mu)-\epsilon)^2 \ \ ,
\end{equation}
where $h_1$ and $h_2$ are large constants.
($h_1 = 10^{12}$ and $h_2 = 10^{15}$ in the calculations presented
here.  Other large values for the constants give similar results.)  
The second and third terms of this expression are penalty functions
that increase the value of $F[\bar{P}(\mu)]$ when constraints (2) and
(3) are not met.

Any standard minimization algorithm can be used.
We performed the minimization using the Splus 
(v. 3.4, Mathsoft, Seattle WA)
gradient-descent function {\it nlminb}.

A minimum may be either global or local.
However, in our problem the minimum is global.
This is guaranteed because the space we are searching ($\bar{P}(\mu) \ge 0$
combined with two linear constraints) is convex, and 
transmitted information is a concave function
with respect to $\bar{P}(\mu)$ \cite[pg.31]{CoverThomas}.  Therefore,
we are guaranteed a single global minimum, and the gradient descent method
must converge to that minimum.

\section*{RESULTS}

We performed new analyses using 
previously published data from 42 V1 complex cells 
from two separate data sets (13 from V1 set 1, 
and 28 from V1 set 2) and
19 inferior temporal (IT) neurons
\cite{bjr90I,Eskandar92I,Kjaer97}.

\subsection*{Log(variance) is linearly related to log(mean).}

\begin{figure}
\epsfig{file=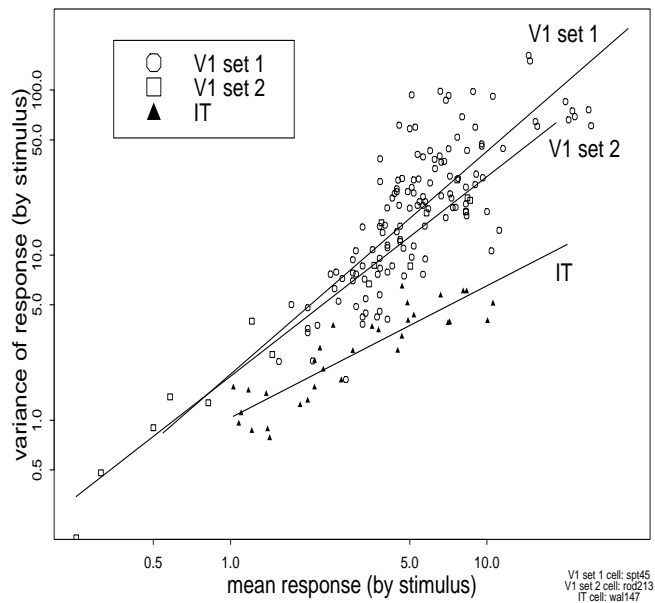,height=3.5 in,width=3.5 in,angle=-90}
\caption{The log(mean) vs. log(variance) regression.
There were 128 stimuli for the V1 set 1 neuron (circles), 16 stimuli 
for the V1 set 2 neuron (squares), and 32 stimuli for the
IT cell (triangles).  The least-squares regression lines for both data
sets are shown.  This example shows the cell with the median
slope from each data set.
}
\label{regression.fig}
\end{figure}

Various researchers have demonstrated
a linear relation between the logarithm of the mean
stimulus-elicited spike count and the logarithm of its variance in
V1 neurons \cite{Tolhurst81,Dean81,Tolhurst83,vanKan85,Vogels89}.
Using linear regression, we find such a relation
for both our V1 complex cells and IT neurons 
(see Figure \ref{regression.fig}).

The slopes of all 13 regressions for neurons in V1 set 1,
in 27 of 28 neurons in V1 set 2, and in
17 of 19 of the IT neurons were 
significant ($p < 0.01$).  
The minimum, median, and maximum values of $r^{2}$ were
0.14, 0.61, and 0.83 in V1 set 1, 0.11, 0.86, and 0.97 in V1 set 2, and
0.02, 0.59, and 0.82 in IT.
The median slopes from the regressions were 1.43 (range 0.91 to 2.67, 12/13 slopes $>$ 1)
and 1.18 (range 0.38 to 1.74, 18/28 slopes $>$ 1) for neurons in V1 set 1 and V1 set 2,
and 0.82 (range 0.41 to 1.53, 14/19 slopes $<$ 1) for IT neurons.  
The median intercepts from the regressions were 0.26 
(range -2.42 to 1.45, 3/13 constants $<$0) and 0.60 
(range -0.79 to 2.10, 5/28 $<$0) in V1 set 1 and V1 set 2, 
and 0.31 (range -1.03 to 1.82, 5/13 $<$0) in IT.

Data arising from a process having equal mean and variance 
(for example, a Poisson process), 
would give rise to a regression intercept and
slope statistically indistinguishable from 0 and
1, respectively.
The regressions from all 13 cells in V1 set 1, 24 of 28 cells in the 
V1 set 2, and 14 of 19 IT cells had either an intercept significantly 
different from 0 ($p<.05$) or a slope significantly different from 1 
($p<.05$), or both. 
These results provide evidence that the Poisson 
distribution does not provide a good model of the data.

\subsection*{Modified Gaussian fits spike count data better than Poisson.}

\begin{figure}
\epsfig{file=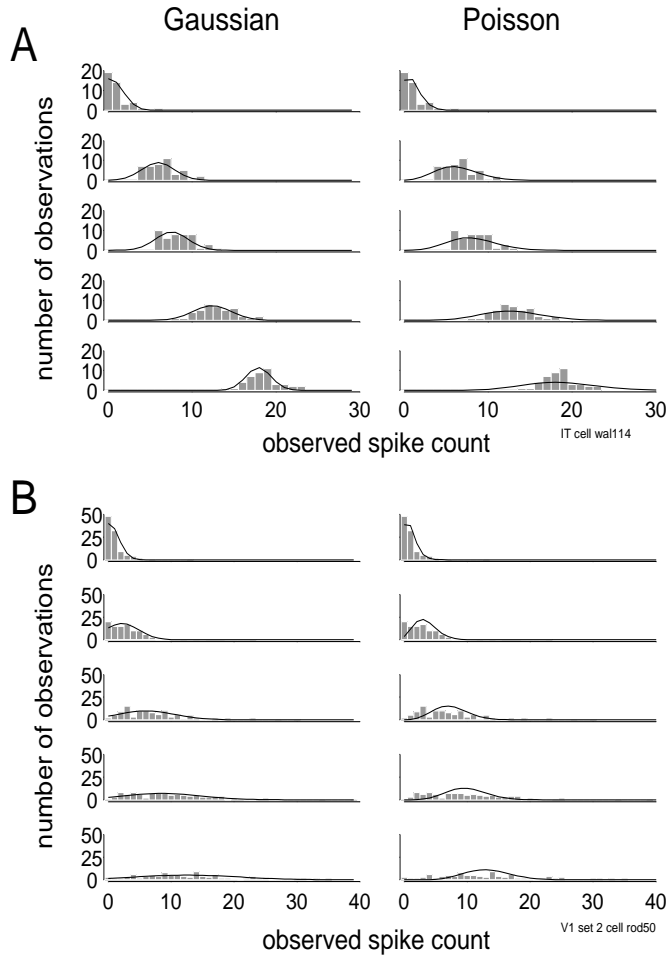,height=5 in,width=3.5 in}
\caption{Sample fits using Poisson and modified Gaussian
distributions.  A:  A cell from IT.  B:  A cell from V1.
Each row shows the histogram of responses to
one of 32 (IT) or 128 (V1) stimuli, 
along with the best-fit modified Gaussian (left)
and Poisson (right) distributions.  The modified Gaussian provides
a better fit, especially when the mean firing rate is large.
The stimuli presented here were selected to show responses with
a range of mean spike counts for each cell.   Note that the scales
for the two sets of graphs are different.
}
\label{distpics.fig}
\end{figure}

We fit modified Gaussian distributions 
(as described in the Methods)  
and Poisson distributions to the empirical distribution of
responses elicited by each stimulus.  
Sample fits are shown in Figure \ref{distpics.fig}.
A $\chi^{2}$
test was used to evaluate the fits.  
The requirement that each response distribution analyzed 
be based on at least twelve presentations of the given stimulus
excluded 7 of 13 of the neurons from V1 set 1.  
Three of 13 cells had enough presentations per stimulus
for all stimuli, and three others had enough presentations for a few stimuli
each, for a total of 433 response distributions.  
All cells from V1 set 2, and all cells from the IT
set, had enough presentations for all stimuli.

The Poisson distribution requires only a single parameter, the mean 
spike count, while the modified Gaussian requires both the mean and 
variance of spike count.
Parameters were computed in three ways:
(1) by choosing the mean (and variance, for the Gaussian)
that minimized $\chi^2$;
(2) by using the observed mean and (for the Gaussian)
the variance predicted by the mean-variance regression;
and (3) by using the observed mean and (for the Gaussian) variance.
Method (3) gave such poor results that we dropped it from
consideration.  
The variance of responses to any given stimulus is a sample variance, and
therefore is itself a random variable.  The regression 
model uses the variances in response to  
all stimuli to estimate the variance of response to each stimulus.
We believe this explains why method (2) is so much more effective than
method (3).

\begin{figure}
\epsfig{file=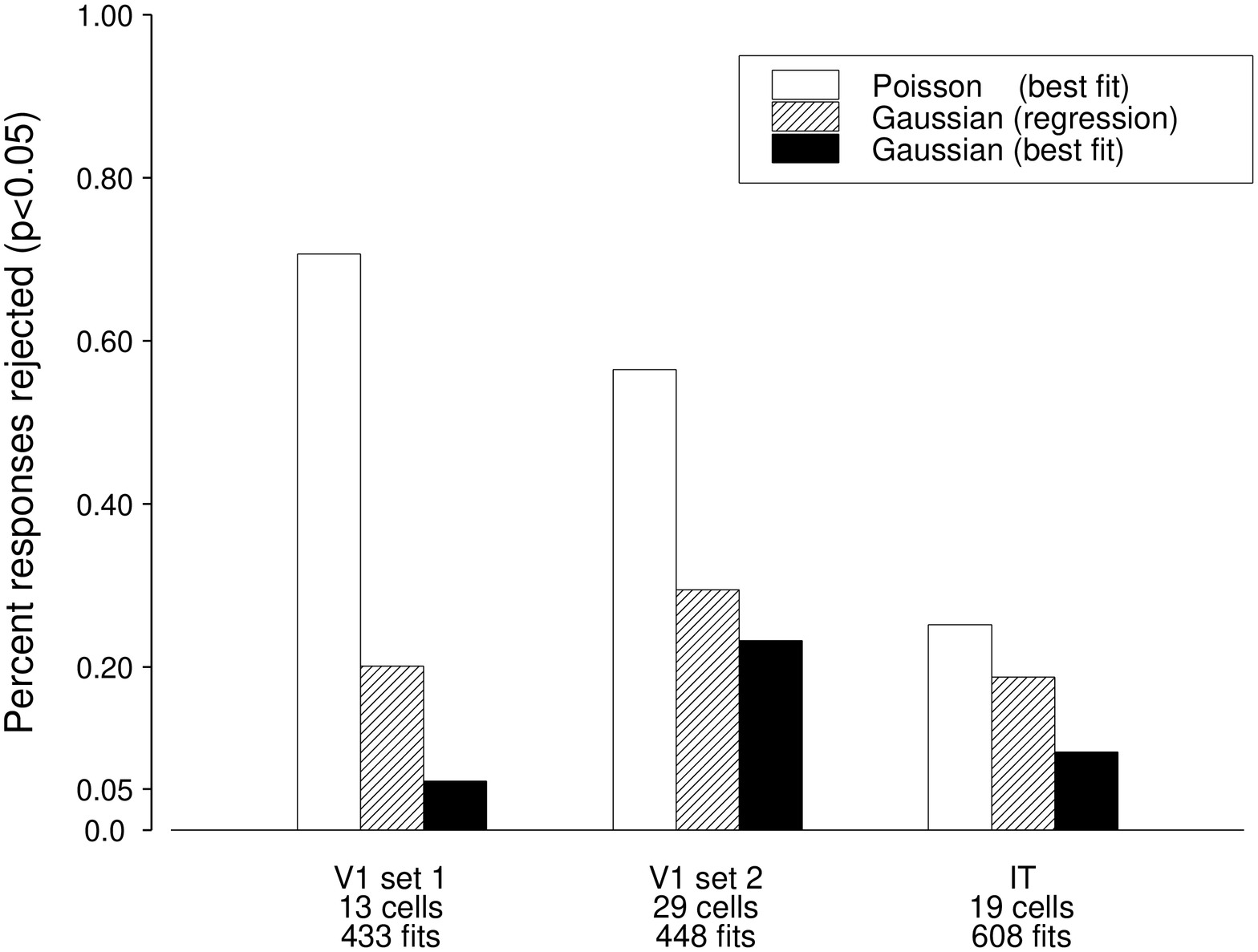,height=3.5 in,width=3 in,angle=-90}
\caption{Chi-squared test of response distributions.
Each bar shows the percent of response distributions
for which the hypothesis that the data came from
the Poisson or modified Gaussian distribution 
can be rejected ($p=0.05$).
The modified Gaussian
distribution using the best-fit
parameters is rejected less often than the distribution using
the observed mean and variance calculated using the log(variance)
vs. log(mean) regression, indicating
that other factors probably influence the variance.
}
\label{chisq.fig}
\end{figure}

Figure \ref{chisq.fig} shows that the Poisson 
distribution could be rejected
($p<0.05$) much more frequently than the
modified Gaussian distribution, even when the best-fit parameters were
used for the Poisson, and the parameters from the Gaussian were estimated
using the data and the mean-variance relation (described below).  
The difference was
even greater when the best-fit parameters were used for the Gaussian as well.

The fact that a $\chi^{2}$ test based on the observed mean and
predicted variance for 
the Gaussian fails more often than 5\% of the time at $p=0.05$
(6\%, 25\%, and 8\% in V1 set 1, V1 set 2, and IT)
suggests that factors other than those
identified in this paper may influence the variance of the distributions.

\subsection*{Information estimates using modified Gaussian distribution.}
 
Because a modified Gaussian distribution modeled the data
better than a Poisson distribution in all three data sets,
we used the modified Gaussian 
to describe the conditional probabilities $P(n|s)$ needed
to compute transmitted information.
We chose
the mean and variance of the modified Gaussian in three ways:
(1) by using the observed
mean together with the variance predicted by the mean-variance
relation; (2) by calculating the mean and variance directly from the data;
and (3) by using the mean and variance
obtained from the fitting procedure.
For comparison we also computed the information using an
artificial neural network 
\cite{Kjaer94,Heller95,Golomb97}.

\begin{figure}
\epsfig{file=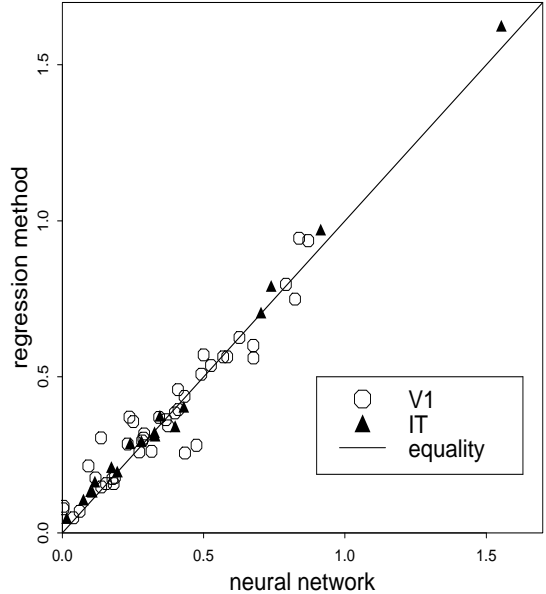,height=4 in,width=3.5 in,angle=270}
\caption{Two methods for estimating transmitted information.
The $x$ axis shows the mean value calculated using the neural
network \cite{Kjaer94}; the $y$ axis shows the value calculated
using the method described in the text.  
The values calculated using the two methods are nearly identical.
All cells with enough data to allow analysis (60) are represented.
}
\label{net.reg.info.fig}
\end{figure}

The estimates obtained using method (1) -- the regression method --
and the network method are nearly equal 
(Figure \ref{net.reg.info.fig}).
Method (2) 
always calculates higher values for transmitted information than
method (1) (mean difference = 0.047 bits, sd = 0.063), and method
(3) calculates even higher values (mean difference = 0.072 bits, sd=0.036).
These represented median percent differences of 8\% and 20\%, respectively. 

As a check, we also calculated the transmitted information on the assumption
that the responses were distributed according to the Poisson distribution.
As expected, given that the Poisson distribution fit the data
poorly, the information estimates showed large deviations from the network
estimates.  The information calculated
using the Poisson distribution was higher than the information
calculated using the modified Gaussian regression method 
in 51 of 60 cells (mean difference = 0.23 bits, sd = 0.25).

\begin{figure}
\epsfig{file=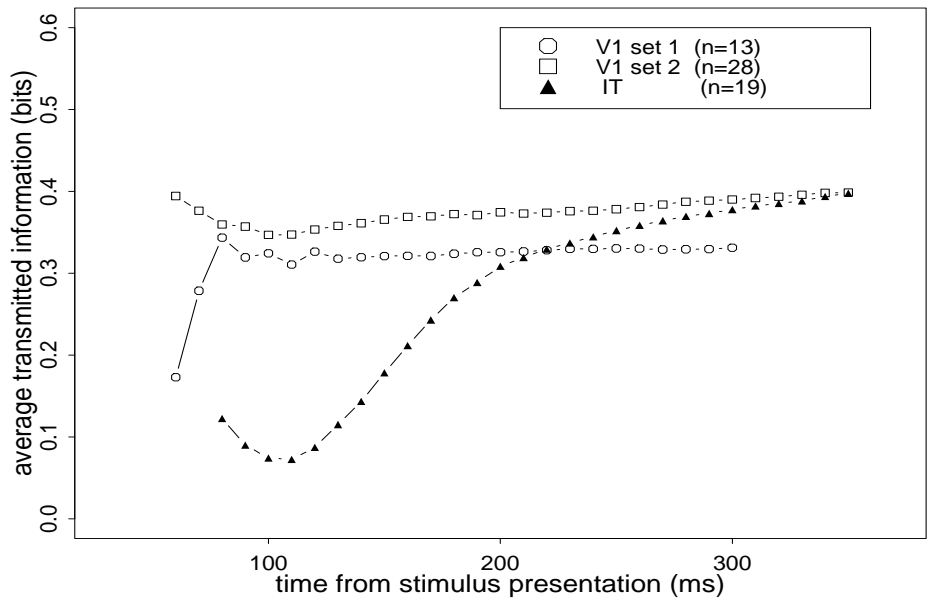,height=5 in,width=3.5 in,angle=270}
\caption{Transmitted information as a function of 
the counting window size.
The $x$ axis shows the time from stimulus presentation.
IT starts later than V1 because it has a longer latency.
The $y$ axis shows the transmitted information accumulated from
stimulus presentation (time 0) to the time indicated on the $x$ axis.
Information accumulates significantly more quickly in neurons from V1
than in neurons from IT.
}
\label{info.window.fig}
\end{figure}

The transmitted information depends on the width of the counting
window.  We examined windows ranging from 30 to 270 ms in V1 set 1, 
from 30 to 320 ms in V1 set 2, and from 50 to 350 ms in IT.  
The log(mean) vs. log(variance) regression was calculated
using the spike count distributions in each window.
The mean information in the
largest time window  was 0.33 bits ($\pm$ 0.16 SD, n=13) for 
V1 set 1, 0.40 bits ($\pm$ 0.25 SD, n=28) for 
V1 set 2, and 0.41 bits ($\pm$ 0.38 SD, n=19) for IT.  
Information rose quickly in the two V1 data sets -- most information
accumulated in just 50 ms (Figure \ref{info.window.fig}).  
Information in IT rose much more slowly,
beginning to level off after approximately 150 ms.
The early dip in transmitted information in cells in IT  
is due to latency effects: some stimuli elicit spikes earlier
than others, and in small windows
this produces information.  Because we are using a spike count code,
information is
reduced as more of the stimuli elicit spikes.  Information rises again as
different spike counts become distinguishable.  This is evidence that
latency carries stimulus-related information in IT neuronal responses.
Latency has been shown to carry stimulus-related 
information in V1 \cite{Gawne96}.

\subsection*{Channel capacity is approximately the same in V1 and IT.}
 
We can compute the channel capacity 
(assuming a spike count code) by finding
the distribution of mean spike counts that yields
the highest transmitted information (see Methods).  
This requires knowing the minimum
and maximum observed spike counts for each cell, and the variability in spike
count at each mean.  The minimum and maximum come directly from the data; 
for the variability
we assumed that the probability of observing a
particular spike count, $P(n|\mu)$, was given by a
modified Gaussian distribution centered on the mean, with a variance
predicted by a linear relation between log(variance) and log(mean).

In the longest windows available,
the minimum spike count for all but 6 cells was 0; that is, at least one
stimulus elicited no spikes.  There were two exceptions in each of the
V1 data sets (in both, the minimum count was 1 spike), and four
exceptions in the IT data set (the minima were 1,2,2, and 4).  The maximum
spikes elicited by any stimulus were fairly evenly spread over a range
of 30-75 in V1 set 1, 15-45 (with two outliers with maxima of  58 and 73) 
in V1 set 2, 
and 10-30 (with two outliers with maxima of 43 and 55) in IT.  
The median spike count maxima were 54, 31, and 24, respectively.

The average channel capacity was 1.26 bits ($\pm$ 0.21 SD) in V1 set 1,
1.12 bits ($\pm$ 0.28 SD) in V1 set 2, and
1.13 bits ($\pm$ 0.47 SD) in IT.  The median channel capacities
were 1.28 bits, 1.02 bits, and 1.23 bits, respectively.

\begin{figure}
\epsfig{file=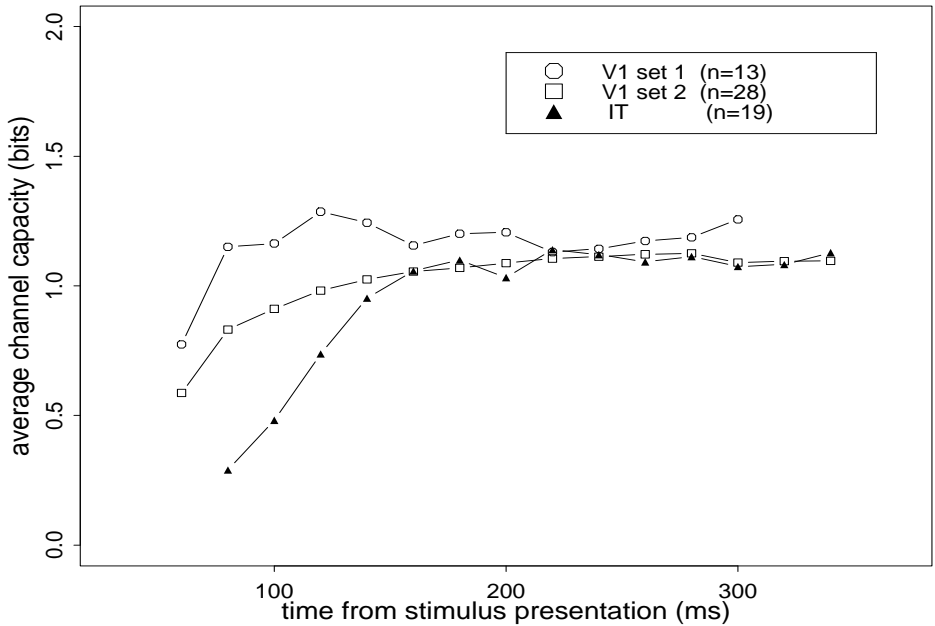,height=5 in,width=3.5 in,angle=270}
\caption{Channel capacity as a function of
the counting window size.
The $x$ axis shows the time from stimulus presentation.
IT starts later than V1 because it has a longer latency.
The $y$ axis shows the channel capacity accumulated from
stimulus presentation (time 0) to the time indicated on the $x$ axis.
Channel capacity rises more quickly in V1 than in IT, although
the difference is not as pronounced as for transmitted information.
}
\label{cc.window.fig}
\end{figure}

As can be seen in Figure \ref{cc.window.fig}, 
channel capacity also rose more quickly as a function of time in
neurons from the two V1 data sets than in neurons from IT, although the
difference is not as pronounced as for transmitted information.

Allowing a larger range of responses increases the amount of information
that can be transmitted.  Therefore, the channel capacity calculated here
depends on the constraints imposed on the dynamic range.
To test the robustness of our numerical results, in a few cases we decreased
$\epsilon$ (which controls how many responses
can lie outside the observed dynamic range; see Methods) 
by a factor of 10 or used a constant instead of a quadratic
weighting function ($C_+(n) = C_-(n) = $ constant; see Methods).
This did not
change the resulting value of the channel capacity by more than 5\% for
any of the examples we considered.
In addition, numerical implementation of the gradient descent 
requires that we
discretize the probability distribution of the mean spike count.  In our
simulations we used a bin size of one spike count, so the means took on
integer values.  Again, to test robustness, in several cases we
decreased the bin size by a factor of 2 and saw little change.

\begin{figure}
\epsfig{file=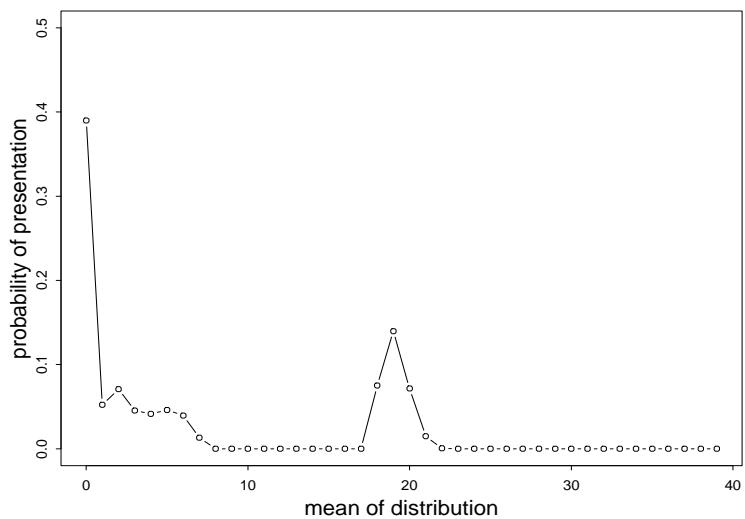,height=4 in,width=3 in, angle=270}
\caption{Distribution of mean responses (each corresponding to a stimulus equivalence
class) that maximizes transmitted information.
The $x$ axis shows the means.  
The $y$ axis shows the probability with which the means
should occur to achieve channel capacity.
The distribution here was calculated using integer means.  As
noted in the text, using a finer grid does not materially affect
the results.
}
\label{cc.probs.fig}
\end{figure}

Figure \ref{cc.probs.fig} shows a typical example of the probabilities
with which various means should occur to achieve
channel capacity.  In every neuron, mean zero (that is, no spikes) 
occurs most frequently.
A small group of means occurs
somewhat less frequently, and the rest of the  means occur with
extremely low probability.  
The bumps appear for different non-zero means 
for different cells; we do not present an average distribution because
averaging  obscures the fact that each distribution consists of discrete
bumps.
If the dynamic range of the cell
is larger, then additional ``ripples''  may appear, indicating further
means that occur with significant probability. 

\section*{DISCUSSION}

In the Introduction we posed two questions:
(1) in what way does the statistical structure of responses
differ across areas?  
and (2) 
how do the differences, if any, affect information transmission?
We found that the responses of neurons in V1 and IT cortex do indeed
have different structures.  The maximum spike count observed in
V1 cortex neurons is generally much higher than that in IT cortex
neurons.  In addition, responses in V1 are much more variable than
those in IT.

Despite the differences in response structure, neurons in V1 and
IT cortex carried approximately the same amount of information about
the stimulus set used in these experiments.
However, transmitted information depends both on the stimuli chosen and on
the relative frequencies with which they are presented.
With a different stimulus set, we might well have measured a different
amount of information.
This makes comparison of transmitted information across areas and
generalization to other stimuli difficult.

To overcome this limitation, we estimated the channel capacity of
the neurons.  The channel capacity is the maximum information that
a neuron can transmit using a given code, given the
constraints of noise in the channel and limited range of responses.
The channel capacities were also approximately equal in the two areas.
This suggests that, in the course of visual processing,
variability is traded off against dynamic range.

The observed differences in spike count and variability are easy both to
visualize and to quantify.  Transmitted information and channel capacity,
on the other hand, are more abstract quantities, and their computation
requires a number of assumptions.  
These assumptions include that spike count is the neural
code, that the observed dynamic range is a good estimate of the true dynamic
range of the cell,
and that the mean-variance relations derived from
our data hold for all visual stimuli.
We now discuss these and other assumptions,
and how they influenced our conclusions.

\subsection*{Transmitted Information}

Given constraints such as finite data sets and limited computational
power, computation of the transmitted information between neural responses
and a stimulus set requires that we choose a neural code.
Here we chose the number of spikes in
a window up to 330 ms wide.  In the two areas we examined, V1 and IT,
such a spike count code has been shown to carry about 80\% of the
information contained in the full neuronal response \cite{Heller95}.
Thus, the true transmitted information is about 25\% higher than the values
we report.  Because we are comparing areas in which the downward bias
caused by using an incomplete code is about the same, this bias has
virtually no effect on our conclusion that the two areas transmit about
the same amount of information.

The primary effect of choosing a spike count code was that it allowed us
to formulate a simple model for the response distributions.  Those
distributions turned out
to be well approximated by a modified Gaussian (see Methods for a precise
definition of ``modified").  The existence of a
model for the response distribution allowed
analyses that would have been impossible otherwise.
Although in principle it is possible to construct a model for more
complicated codes, it is more difficult and larger data sets may be
required.

\subsection*{Channel Capacity}

An intrinsic drawback of transmitted information is that it depends on
the frequencies with which stimuli are presented.  This makes the value
of the transmitted information somewhat arbitrary | it can almost always
be made either larger or smaller simply by changing the probabilities
with which stimuli are presented.  One could imagine adjusting
stimuli probabilites to maximize the transmitted information.
Shannon and Weaver (1949)
defined the resulting maximum value as the channel capacity.  It
is a function only of the conditional probability distribution $P(r|s)$.

The use of the term channel capacity to represent the 
maximum amount of information
that can be transmitted using a given code (or ``alphabet'', in 
Shannon's original terminology) in the presence of noise is well
established \cite{CoverThomas,ShannonWeaver}.  
Channel capacity, like transmitted information, 
depends on how we choose to interpret the cell's response,
that is, on our assumption about the neural code.   
However, once we choose a code, the channel capacity
is well-defined.  Since it is always possible that some other code would allow
the cell to transmit more information than the code under examination,
the channel capacity based on any given code is a lower bound on the amount 
of information that the cell can transmit. 
Because the spike-count code has been shown to carry 
about 80\% of the stimulus-related
information \cite{Heller95}, it provides a reasonable
first approximation.

The actual code used by neurons is likely to
include some temporal aspects of the response.  
Although temporal modulation could provide many degrees of freedom,
studies of V1 neurons have
shown that only a few degrees of freedom are used to
carry stimulus-related information \cite{bjr90I,Heller95,Victor96}.
We predict that the increase in channel capacity when temporal
modulation is taken into account will be proportional to the
increase in transmitted 
information.  If this is true, then the actual channel capacities
will be about 25\% larger than the values we calculated.

Channel capacity
should not be confused with information capacity
of the signal \cite{MacKayMcCulloch}.  
The information capacity is the information present in 
the signal itself, subject to a model of the noise.

\subsection*{Calculating Channel Capacity}

To calculate transmitted information 
we need an estimate for the response
distribution for each stimulus presented in an experiment.  
These distributions can be estimated directly from the data.
Calculating channel capacity is more difficult, because it 
requires knowing the response distribution for all stimuli, not
only those presented in a particular experiment.
This problem can be overcome by 
sorting stimuli into groups based
on the response distribution each evokes.
Here we will call each such group of stimuli an {\it equivalence class}.  
The neuron cannot distinguish members of an equivalence
class from one another.  
For example, otherwise identical stimuli of different colors
produce the same response distribution in a cell insensitive to color.  
Therefore, rather than considering
each stimulus separately, we work with the equivalence classes.

There are likely to be equivalence classes we do not observe experimentally.
However, if we can describe the set of equivalence classes with a model
involving a small number of parameters, and if we can
show (as we have in the Results)
that the model adequately describes the distributions at parameter
values observed experimentally, then
we can assume that the model also describes the distributions for
unobserved parameter values.  This overcomes the major obstacle to 
calculating channel capacity.

We found that a modified Gaussian distribution provides a good model 
of the response distributions in our experiments.  
We compared transmitted information values obtained using the
modified Gaussian to the values obtained by a previously
validated method using a neural network 
\cite{Kjaer94,Heller95,Golomb97}; 
the values obtained by the two
methods are indistinguishable.  Thus, although there may be a distribution
that fits these data better than the modified Gaussian, the modified
Gaussian is a good model for the calculations we want to perform.
A Poisson distribution, although
often used to model responses, fit our experimental data poorly.
Others have 
reached the same conclusion \cite{Softky93,Victor96}.

The modified Gaussian distribution is fully specified
by two parameters:  the mean and variance.  
This distribution provides a sufficiently simple model of the
neuronal responses to allow calculation of the channel capacity.
It turns out that we can simplify the problem by estimating the
variance of each distribution based on the mean. 
This simplification is achieved using the linear relation between
log(mean) and log(variance)
\cite{Tolhurst81,Dean81,Tolhurst83,vanKan85,Vogels89}. 
If we know the mean of a response, we can calculate its variance. 
Therefore, any response distribution can be characterized by its mean.

With this model,
the equivalence classes are labeled by the mean response.
Two stimuli that produce the same mean response also
produce identical response distributions.
These stimuli will be indistinguishable based on the responses of
the cell, and need not be considered separately.  
Note that every stimulus produces {\it some} mean spike count,
even if it is zero, and therefore is accounted for in this model.

We now invoke our main assumption:  that mean responses not observed
in our experiments could be observed given appropriate stimuli.
Then the channel capacity can be calculated by finding
the distribution of mean spike counts
that maximizes transmitted information.  

It is, of course, necessary to restrict the range of spike counts.
There are both biophysical and mathematical reasons for this.
Biophysically, we know that all cells have a maximum
firing rate.  Mathematically, if we allow an infinite number
of spikes, the channel capacity would be infinite.
We restricted the
range of spike counts to be consistent with observed data.
Briefly, we required that the spike count fall primarily within the
experimentally observed minimum and maximum (see Methods).
The maximum experimentally observed spike count may be an
underestimate of the true maximum,
as we may not have used stimuli
that elicited the highest firing rates.  
However, the peak firing
rates that we saw in V1 and IT are similar to those
seen by others \cite{Tolhurst81,Tolhurst83,Vogels89,Rolls82,Rolls84,Perrett84}.
If new evidence does show that the dynamic range is larger than we
observed, the channel capacity can be recalculated.
The effects, however,
are modest.  
We calculated the increase in channel capacity
assuming that the maximum firing rate for each
cell was 25\% greater than the measured values.  The median
increase in channel capacity was 8.5\% (range 0.8\% to 20.9\%).

To ensure that the estimate of channel capacity is reasonable, it
is important to know that the log(mean) vs. log(variance) regression
is reliable.  For all but 3 of 60 neurons the regressions were significant.
It is possible that the regression could change for different
stimuli, or, for example, in different attentional states. 
A change in the regression would provide powerful evidence for a state
change at a fundamental level of neural function.
Such changes have been reported for fly H1 cell \cite{dRvS97}.
de Ruyter van Steveninck et al. (1997) 
found less variance, and more information, in the
responses of fly H1 cell to a moving coherent stimulus when the stimulus
moved along a ``presumably more naturalistic'' two-dimensional trajectory
than when the stimulus moved in one direction at constant speed.   
However, at least
one study that looked for such differences in one monkey visual
cortical area (MT) failed to find them \cite{mcadams96}.

When we calculated the channel capacity we found that, on average,
it is about the
same in V1 and IT, and in both areas it is about 2-4 times 
the transmitted information.  This lends support to the notion that
the smaller dynamic range found
in IT (as compared to V1), which would tend to decrease the information
that can be transmitted, is balanced by less variable neuronal responses,
which tend to increase information transmission.

\subsection*{Comparison with other studies}

In this study the information transmitted by the spike count
averaged about 1 bit/300 ms (3 bits/sec).  The channel
capacity, while typically 2-4 times larger in any cell,
is also not very large.  Other investigators have reported
significantly higher transmission rates.  
A recent preliminary report \cite{buracasetal97} indicates that
the transmitted information bit rates of MT neurons in the monkey reach
30 bits/sec with moving stimuli.  de Ruyter van Steveninck et al. (1997)
estimated that the responses of the H1 neuron of the fly contain
2.43 bits/30 ms (about 80 bits/sec).
Here we examine factors
that may account for the differences in our results.

Both MT cortex and the fly H1 neuron analyze motion.
In the experiments noted above, monkeys or flies were
shown stepped motions of coherent patterns of bars.
Analysis was carried out to determine the information transmitted
by the neurons about the direction of motion of the (unchanging)
coherent pattern.  Motion analysis (especially in only one or two dimensions) 
is an easier problem than pattern recognition. 
We expect that it requires less computation than pattern recognition,
resulting in increased
information transmission rate per neuron.

For the neurons in our experiments, almost
all of the stimulus-related information 
that is available in the spike count 
is available in the first 50
ms of information transmission (after a 30 ms latency period) in V1
cortex, and in the first 200 ms of information transmission (after
a 50 ms latency period) in IT cortex.  If these peak information
rates were maintained, the V1 and IT neurons would be able to transmit
approximately 20 and 5 bits/sec, respectively.  These rates are still
significantly smaller than those reported in the motion studies, although
20 bits/sec approaches the rates seen by Buracas et al. (1996) in MT.

Given that, in V1, most information that will ever be available is available
within 50 ms of the beginning of a response, why flash stimuli 
on a screen for 300 ms?
During normal primate vision, a new image appears on
each receptive field 1-3 times/sec due to saccadic eye movement,
after which the image is kept nearly still on the retina (compared
to saccade velocities).  
Therefore, to study the processes underlying pattern recognition, flashing
stimuli onto the visual field at relatively slow rates seems an
appropriate paradigm.  It remains to be seen whether more rapid presentation
of the images would allow consistent peak information transmission, or
whether the images would interfere with one another.

In both of the analyses of motion, temporal aspects of neural response
were taken into consideration.  In our analysis, they were not.
Previous analyses \cite{Heller95} found that spike count transmitted
approximately 80\% of the information available in the full
response.  Therefore, if we accounted for temporal aspects of the signal,
we could expect a 25\% rise in transmitted information.

\subsection*{Questions Raised}

Presumably neurons in these two regions operate
according to the same biophysical principles.  How is it that
the variance is lower in IT neurons than in V1 neurons?  Does the larger
dynamic range with larger variance offer some advantage
that offsets the energy cost of higher firing rates?
Finally, why don't all neurons use a large dynamic range with low
variability?

\section*{Acknowledgements}

The authors thank Dr. Mike W. Oram and Dr. Karen D. Pettigrew
for helpful discussion and comments on the manuscript.

\end{spacing}

\footnotesize
\begin{thebibliography}{12345678901234567 19nnx}

\renewcommand{\em}{}

\bibitem[Buracas et al., 1996]{buracasetal97}
Buracas, G., Zador, A., DeWeese, M., Albright, T. (1996)
Measurements of information rates in monkey MT neurons in
response to time-varying stimuli.
{\it Society for Neuroscience Abstracts} 22: 717.

\bibitem[Cover and Thomas, 1991]{CoverThomas}
Cover, T.M. and Thomas, J.A.
(1991)
Elements of Information Theory,
Wiley \& Sons, New York.

\bibitem[de Ruyter van Steveninck et al., 1997]{dRvS97}
de Ruyter van Steveninck, R.R., Lewen, G.D., Strong, S.P.,
Koberle, R., and Bialek, W.
Reporducibility and variability in neural spike trains.
{\it Science} 275: 1805-1808.

\bibitem[Dean, 1981]{Dean81}
Dean, A.F.
(1981)
The variability of discharge of simple cells in the cat striate cortex.
{\it Exp Brain Res} 44:437-40.

\bibitem[Eskandar et al., 1992]{Eskandar92I}
Eskandar, E.N., Richmond, B.J. and  Optican, L.M.
(1992)
Role of inferior temporal neurons in visual memory. 
I. Temporal encoding of information about visual images, 
recalled images, and behavioral context.
{\it J Neurophysiol}  68: 1277-95.

\bibitem[Foldiak 1993]{Foldiak93}
Foldiak, P. (1993) The 'ideal homunculus':  Statistical inference
from neural population responses.  In (F.H. Eeckman \& J.M. Bower,
Eds.), {\it Computation and Neural Systems}, pp. 55-60.  Kluwer
Academic Publishers, Norwell, MA.

\bibitem[Gawne et al., 1996]{Gawne96}
Gawne, T.J., Kjaer, T.W. and Richmond, B.J. (1996) Latency:
Another potential code for feature binding in striate cortex.
{\it J. Neurophysiol.} 76: 1356-1360.

\bibitem[Gershon et al., 1996]{Gershon96}
Gershon, E.D., Latham, P.E., Jin, G-X and Richmond, B.J.
(1996)
Stimulus-elicited neuronal responses in striate and inferior temporal
cortices are well described by a Gaussian distribution.
{Soc. Neurosci. Abs.} 22: 1612.

\bibitem[Golomb et al., 1997]{Golomb97}
Golomb, D., Hertz, J., Panzeri, S., Treves, A., and Richmond, B.
(1997) How well can we estimate the information carried in neuronal
responses from limited samples?
{\it Neural Computation} 9: 649-665.

\bibitem[Heller et al., 1995]{Heller95}
Heller, J., Hertz, J.A., Kjaer, T.W. and Richmond, B.J. (1995)
Information flow and temporal coding in primate pattern vision.
{\it J. Computational Neurosci.}
2: 175-193.

\bibitem[Kendall and Stuart, 1961]{KendallStuart}
Kendall, M.G., and Stuart, A. (1961)
{\it The Advanced Theory of Statistics, vol. 2:  
Inference and Relationship.}
Hafner, London.

\bibitem[Kjaer et al., 1994]{Kjaer94}
Kjaer, T.W., Hertz, J.A. and Richmond, B.J.
(1994)
Decoding Cortical Neuronal Signals: Network Models,
Information Estimation and Spatial Tuning.
{\it J Computational Neuroscience}
1: 109-139.

\bibitem[Kjaer et al., 1997]{Kjaer97}
Kjaer, T.W., Gawne, T.J., Hertz,J.A.
and Richmond, B.J. (in press) Insensitivity of V1 complex cells to
small shifts in the retinal image of complex patterns.  {\it J.
Neurophysiol.}

\bibitem[MacKay and McCulloch, 1952]{MacKayMcCulloch}
MacKay, D.M. and McCulloch, W.S. (1952)
The limiting information capacity of a neuronal link.
{\it Bulletin Math. Biophysics} 14: 127-135.

\bibitem[McAdams and Maunsell, 1996]{mcadams96}
McAdams, C.J. and Maunsell, J.H.R. (1996)
Attention enhances neuronal responses without
altering orientation selectivity in macaque area V4.
{\it Society for Neuroscience Abstracts} 22: 1197.

\bibitem[Perrett et al., 1984]{Perrett84}
Perrett, D.I., Smith, P.A.J., Potter, D.D., Mistlin, A.J.,
Milner, A.D. and Jeeves, M.A.
(1984)
Neurones responsive to faces in the temporal cortex: studies
of the functional organization, sensitivity to identity and relation
to perception.
{\it Human Neurobiol.} 3: 197-208.


\bibitem[Richmond et al., 1990]{bjr90I}
Richmond, B.J., Optican, L.M., and Spitzer, H. (1990)
Temporal encoding of two-dimensional patterns by single units in
primate visual cortex
I. Stimulus-response Relations.
{\em J Neurophysiol}
64: 351-369.

\bibitem[Rolls et al., 1982]{Rolls82}
Rolls, E.T., Perrett, D.I., Caan, A.W. and Wilson, F.A.W.
(1982)
Neuronal responses realted to visual recognition.
{\it Brain} 105: 611-646.

\bibitem[Rolls, 1984]{Rolls84}
Rolls, E.T.
(1984)
Neurons in the cortex of the temporal lobe and in the
amygdala of the monkey with responses selective for faces.
{\it Human Neurobiol.} 3: 209-222.

\bibitem[Shannon and Weaver, 1949]{ShannonWeaver}
Shannon, C.E., and Weaver, W. (1949)
{\it The mathematical theory of communication.}
U. Illinois Press (Urbana), 1949.

\bibitem[Softky and Koch, 1993]{Softky93}
Softky, W.R. and Koch, C.
(1993)
The highly irregular firing of cortical cells is inconsistent
with temporal integration of random epsps.
{\it J. Neurosci.} 13: 334-350.

\bibitem[Tolhurst et al., 1981]{Tolhurst81}
Tolhurst, D.J., Movshon, J.A. and Thompson, I.D.
(1981)
The dependence of response amplitude and variance of
cat visual cortical neurones on stimulus contrast.
{\it Exp. Brain Res.}
41: 414-419.

\bibitem[Tolhurst et al., 1983]{Tolhurst83}
Tolhurst, D.J., Movshon, J.A. and Dean, A.F.
(1983)
The statistical reliability of signals in single neurons in cat
and monkey visual cortex.
{\it Vision Res.}
23: 775-785.

\bibitem[van Kan et al., 1985]{vanKan85}
van Kan, P.L.E., Scobey, R.P. and Gabor, A.J.
(1985)
Response covariance in cat visual cortex.
{\it Exp. Brain Res.} 60: 559-563.

\bibitem[Victor and Purpura, 1996]{Victor96}
Victor, J.D. and Purpura, K.P.
(1996)
Nature and precision of temporal coding in visual cortex:
A metric-space analysis.
{\it J. Neurophysiol.} 76: 1310-1326.

\bibitem[Vogels et al., 1989]{Vogels89}
Vogels, R., Spileers, W. and Orban, G.A.
(1989)
The response variability of striate cortical neurons in the behaving
monkey.
{\it Exp. Brain Res.} 77: 432-436.

\end{thebibliography}
\end{document}